# Design and Development of a Novel Soft and Inflatable Tactile Sensing Balloon for Early Diagnosis of Colorectal Cancer Polyps

Ozdemir Can Kara*, Hansoul Kim*, Jiaqi Xue, Tarunraj G. Mohanraj, Yuki Hirata, Naruhiko Ikoma, and Farshid Alambeigi

*Abstract*— In this paper, with the goal of addressing the high early-detection miss rate of colorectal cancer (CRC) polyps during a colonoscopy procedure, we propose the design and fabrication of a unique inflatable vision-based tactile sensing balloon (VTSB). The proposed soft VTSB can readily be integrated with the existing colonoscopes and provide a radiation-free, safe, and high-resolution textural mapping and morphology characterization of CRC polyps. The performance of the proposed VTSB has been thoroughly characterized and evaluated on four different types of additively manufactured CRC polyp phantoms with three different stiffness levels. Additionally, we integrated the VTSB with a colonoscope and successfully performed a simulated colonoscopic procedure inside a tube with a few CRC polyp phantoms attached to its internal surface.

## I. INTRODUCTION

The prevalence of colorectal cancer (CRC) has rapidly increased worldwide in recent years [1]. Based on the population growth and aging estimates, new cases of CRC are projected to reach 3.2 million by 2040 [2]. On the other hand, a high survival rate (91%) of the patients can be achieved if CRC is promptly diagnosed in an early stage [3]. In other words, since survival outcomes differ significantly based on the tumor stage at the time of detection, early detection of pre-cancerous lesions (i.e., polyps) via colonoscopy has a significant impact on treatment outcomes and is of paramount importance [4].

Colonoscopy is the gold standard for the evaluation of CRC due to its ability to visualize the inner surface of the colon (where cancers arise), acquire biopsies, and treat pre-cancerous as well as early-stage cancers. Nevertheless, some types of polyps have a flat geometry (98% of missed polyps [5]) with small size, slight elevation/depression, and minor textural details in the early stages making their visual detection and classification using traditional colonoscopy an arduous procedure for the clinicians [6]. This has resulted

Research reported in this publication was supported by the National Cancer Institute of the National Institutes of Health under Award Number R21CA280747 and Proof of Concept Award in The University of Texas at Austin.
*Authors contributed equally to this work.
Ozdemir Can Kara, Hansoul Kim, Tarunraj G. Mohanraj, Jiaqi Xue and Farshid Alambeigi are with the Walker Department of Mechanical Engineering, University of Texas at Austin, TX, USA Email: {ozdemirckara,jiaqixue, tarunrajgm}@utexas.edu, and {han.kim,farshid.alambeigi}@austin.utexas.edu
Yuki Hirata and Naruhiko Ikoma are with the Department of Surgical Oncology, Division of Surgery, The University of Texas MD Anderson Cancer Center, Houston, TX, USA, 77030. Email: {yhirata,nikoma}@mdanderson.org

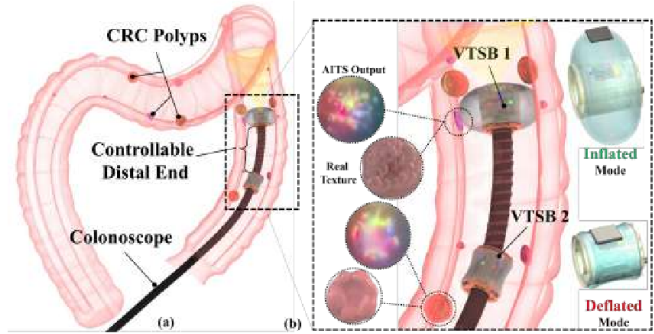

Fig. 1. (a) Conceptual illustration of integrating two VTSBs with a colonoscope for CRC polyp detection. (b) A close-up view of inflated VTSB 1 and deflated VTSB 2 together with their potential textural image outputs.

in the early detection miss rate (EDMR) of approximately 26% for adenomas (non-cancerous polyps), 27% for serrated polyps (saw-toothed appearance under the microscope), and 34% for flat adenomas during conventional colonoscopic procedures [7]. To address these challenges, sophisticated imaging modalities (e.g., narrow-band imaging (NBI)) have also been developed [8], [9]. Despite their benefits, these approaches and particularly chromo-endoscopy have a slow learning curve requiring over 200 procedures for an endoscopist to be considered competent in the technique [6]. This demands the development of new diagnosis devices to provide high-resolution textural information to minimize the current high EDMR of CRC polyps.

Recent literature reports that types and stages of CRC polyps can be determined through their morphology including the size, shape, and textural details of a polyp [10], [11]. However, the task of polyps detection and classification using their morphology is highly complex and clinician-dependent [12], [13], therefore, increasing the risk of EDMR and mortality. Of note, this can be mainly attributed to the limitations of existing vision-based colonoscopic technologies (e.g., camera occlusion and blur), lack of haptic feedback, and polyps classification approaches, causing early diagnosis of CRC to be very challenging for clinicians [14], [15]. Therefore, morphological characteristics have not yet fully been integrated to guide treatment strategies.

To address the aforementioned critical limitations in the early diagnosis, and classification of polyps, several endoscopic instruments with embedded tactile sensors have been developed and reported in the literature. For instance, a piezoelectric tactile sensor embedded in rigid endoscopic instruments has been presented by Chuang et al. [16]. Despite its benefits, this technology cannot provide surface texture under static conditions and must be actuated at a high frequency with a shaker. In [17], a micro electromechanical-based endoscopic tactile sensor in a layered structure has also been developed, however, fabrication of this sensor is relatively burdensome, and cannot provide the surface texture of the polyp. Ly et al. [18] also presented a palpation system consisting of a pneumatic tactile display and a force sensor to perceive the tumor size. The reported accuracy of this sensor is 65% for the size which may not be sufficient for a sensitive and reliable early diagnosis of cancer polyps. In another effort, distributed electrode arrays were also employed to measure the interaction force within a relatively large potential contact area, still, the creation of surface textures is complex [19]. A finger-type tactile sensor has also been developed by Arian et al. [20] for polyp palpation. However, this sensor requires a large interaction force, which may damage the soft tissue during diagnosis. Recently, we successfully investigated the potential of utilizing a large vision-based surface tactile sensor for early diagnosis of CRC polyps outside of the patient's body [21]. Despite the high-resolution textural output of this sensor and constructing rigid elements and structure of this class of tactile sensors make it almost impossible to miniaturize such a device and utilize it for CRC diagnosis through colonoscopy. A detailed review of similar technologies can also be found in [22].

As our main contributions of this study, to collectively address the above-mentioned requirements and limitations of current studies and in order to obtain a radiation-free, safe, and high-resolution textural mapping and morphology characterization of CRC polyps, we propose a unique inflatable vision-based tactile sensing balloon (VTSB). As shown in Fig. 1, the proposed soft VTSB (i) provides a safe and high-resolution textural mapping of CRC polyps that can be used for detection and classification of type and stage of identified cancer polyps [21]. Of note, current topographic mapping technologies are all developed solely based on visual colonoscopy images (e.g., NeoGuide Endoscopy system [23]), and (ii) readily integrates with the current colonoscopic devices and robots to provide clinicians with additional diagnostic information while not changing their current clinical procedures. The performance of the proposed VTSB has been thoroughly characterized and evaluated on 4 different types of CRC polyp phantoms with three different stiffness levels at different interaction force and VTSB pressure values. Moreover, to evaluate the efficacy of the VTSB, we integrated the VTSB with a colonoscope and performed a simulated colonoscopic procedure inside a tube with a few CRC polyp phantoms attached on its internal surface.

## II. Methodology

### A. Working Principle and Diagnosis Procedure

As shown in Fig. 2, the proposed VTSB design is composed of a ring-shape frame housing a camera, a stretchable torus-shape balloon that is fixed to the frame and can be inflated or deflated using the air pressure lines connected to the ring-shape frame, and small LEDs for illuminating the inflated region inside the balloon and provide adequate lighting condition for the cameras. As demonstrated in Fig. 1, the *working principle* of VTSB is very simple yet highly intuitive in which the deformation caused by the interaction of the soft inflated VTSB's balloon layer with the colon surface can visually be captured by the embedded camera.

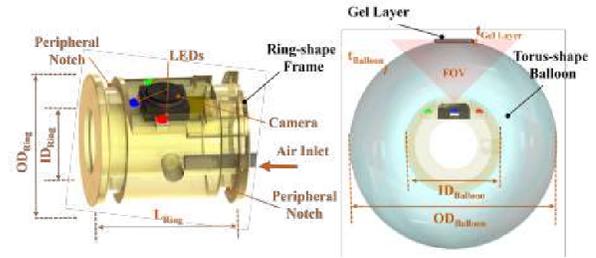

Fig. 2. Proposed design for the Ring-shape Frame and Torus-shape balloon.

In the *envisioned diagnosis procedure*, as conceptually shown in Fig. 1, first, one or multiple of the VTSBs (in their deflation mode) will be integrated with the flexible tubes of the colonoscope. Next, following the conventional screening procedure, as the clinician inserts and advances the colonoscope in the colon, he/she sequentially inflates VTSBs through the process to obtain a high-resolution image of the section of the colon, ensuring a safe interaction force (i.e., <13.5 N [24]), and a safe pressure level (i.e., <7.6 kPa [25]). Of note, the VTSB attached to the distal end of the colonoscope (i.e., VTSB1 in Fig. 1) can also be actuated with colonoscope for further diagnosis and direct interaction and screening of specific areas within the colon. Notably, similar to many FDA-approved colonoscopic rings and cuffs listed in [26], integrated VTSBs will not change the existing insertion procedure and, therefore, we do not expect to create any safety concerns.

### B. Design of the VTSB

As shown in Fig. 2, the proposed VTSB design is composed of the following main components:

*1) Ring-Shape Frame:* This frame houses (1) a camera, (2) a torus-shape balloon that is fixated to the frame, (3) small LEDs for illuminating the inflated region inside the balloon, and (4) air pressure inlets and outlets for inflation and deflation of the balloon. The inner ($ID_{Ring}$) and outer ($OD_{Ring}$) diameters of the frame are determined based on the outer diameter of the colonoscope (typically ranging between 12-15 mm) and the size of the colon diameter (i.e., ranging between 3 to 8 cm [27]). Also, ($L_{Ring}$) of the ring frame determines the diagnostic area of the colon and is designed based on the size of the balloon, embedded cameras, and LEDs. Moreover, this frame has two peripheral notches for fixating the balloon and avoiding the air leak. Based on these design criteria and our utilized PENTAX EC 3840 L colonoscope (shown in Fig. 4) with 13 mm OD, we selected (i) $ID_{Ring}$= 13.2 mm, (ii) $OD_{Ring}$= 26.5 mm to avoid blocking the colon and creating a potential injury, and (iii) a 5 MP camera (Arducam 1/4 inch 5 MP sensor min) as well as miniature high power Red, Green, and Blue LEDs with 2.45 mm × 2.45 mm × 1.84 mm size (941-XBDRED0801, 941-XBDGRN0D01, 941- XBDBLU0202). To minimize the overall size of the ring-shape frame, we embed cameras,

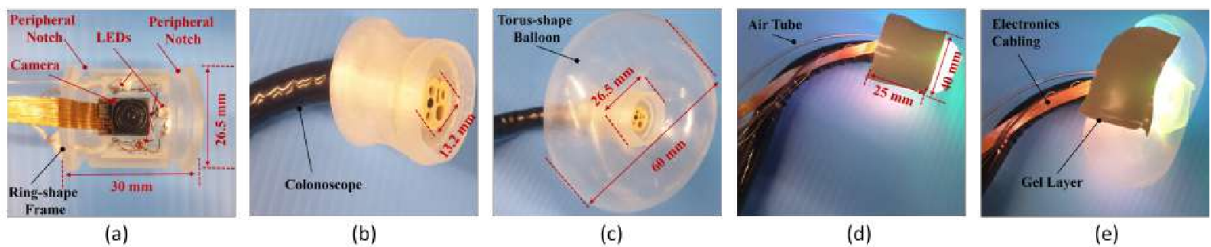

Fig. 3. (a) The fabricated ring-shape frame housing the camera, RGB LEDs, and the pneumatic lines, (b) The deflated mode of the fabricated balloon integrated with the distal end of a colonoscope, (c) The inflated mode of the fabricated balloon integrated with the distal end of a colonoscope, (d) The deflated mode of the VTSB integrated with the distal end of a colonoscope, (e) The inflated mode of the VTSB integrated with the distal end of a colonoscope.

LEDs, and air channels within the frame and fabricated it using high-resolution 3D printing with the Stereolithography (SLA) 3D Printer (Formlabs Form 3, Formlabs Inc). Of note, one of the beneficial features of the VTSB is its compact overall size that solely depends on the size of the ring frame (since the balloon layer does not add to the VTSB size) that readily can be scaled depending on the VTSB constructing components and colonoscope diameter. Fig. 3 (a) shows the fabricated ring-shape frame assembled with the camera, LEDs, and pressure lines.

*2) Stretchable Torus-Shape Balloon:* As the deformable and stretchable interface between the VTSB and CRC polyps, the balloon layer provides a sufficient and safe interaction force for the gel layer of the VTSB to obtain detailed morphology of polyps.

*3) VTSB Gel Layer:* Gel layer is the most essential component of the VTSB that acts as the sensing layer of the proposed tactile sensing device. Of note, the fabrication procedure, stiffness, and thickness of this layer directly affect the sensitivity and performance of the sensor.

### C. Fabrication of the Inflatable Balloon and Gel layer

The following describes a two-step molding procedure for the fabrication of the inflatable balloon and the gel layer in which we first fabricate the inflatable balloon and then directly mold the flexible gel layer on this balloon.

*1) Inflatable Balloon:* To create an inflatable torus-shape balloon for the proposed VTSB, we first designed and additively manufactured a cylindrical-shape mold consisting of three parts (i.e., a hollow tube, an inner circular rod, and a bottom fixture part for the rod) using an FDM 3D printer (Raise 3D Pro2 Series, RAISE3D Inc.) and an Art White PLA filament (RAISE3D Inc.). The inner diameter of the hollow tube and the outer diameter of the inner circular rod were 26 mm and 24 mm, respectively. The overall length of the fabricated mold was 60 mm. Next, to fabricate the inflatable balloon using the manufactured mold, we used a soft water clear biocompatible platinum cure silicone (Ecoflex 00-31 Near Clear, Smooth-On, Inc.) and mixed its two parts, consisting of Part A and Part B, with the ratio 1:1. Before initializing the molding process, the surface of the three-part mold was coated with Ease 200 (Mann Technologies) to prevent adhesion. Next, the silicone mixture was put into a vacuum chamber to remove the bubbles inside it. After the degassing step, the silicone mixture was poured into the three-part mold to fill the empty space between the hollow tube and the inner rod. After a curing process of about 4-5 hours at room temperature, the mold is disassembled to obtain the fabricated durable, soft, and inflatable balloon. Fig. 3 (b)-(c) shows the deflated and inflated modes of the fabricated balloon integrated with the ring-shape frame and the colonoscope.

*2) VTSB Gel Layer:* To precisely map and identify the textural details of the CRC polyps with high sensitivity and resolution, we fabricated a rectangular sensing gel layer (length × width × thickness as 40 mm × 25 mm × 3 mm) from a softer silicone than the one used for fabrication of balloon (i.e., platinum-catalyzed silicones Ecoflex 00-10, Smooth-On, Inc.) directly on the surface of the fabricated balloon. To this end, we first constructed a three-part rectangular mold (80 mm × 80 mm × 21 mm) including one rectangular block for fixating the fabricated balloon and two rectangular parts with a rectangle opening (40 mm × 25 mm × 3 mm) in the middle for pouring of the silicone mixture after degassing process in a vacuum chamber. Of note, the size of this opening determines the size of the gel layer or sensing space of the VTSB. After pouring the degassed silicone mixture into the rectangular opening, and curing for 4 hours at room temperature, the mirror effect nail powder (Silver Chrome Nail Powder, Pretty Diva) was brushed on the gel layer's surface to avoid light leakage and improve the sensitivity of the VTSB. Of note, a mirror effect nail powder was selected because it has been shown that the sensitivity of the VTSB is correlated to the reflectance of the coating material and can be increased by utilizing the reflective layer [28]. Finally, a thin layer of silicone mixture (Ecoflex 00-10) consisting of grey pigment (a blend of both black and white pigments- Silc Pig Black, Silc Pig White, Smooth-On Inc) was poured, on the surface of the VTSB gel layer to prevent light leakage and stabilize the nail powder layer. Fig. 3 (d)-(e) shows the fabricated gel layer attached to the inflatable balloon and mounted at the distal end of a colonoscope.

### III. EVALUATION EXPERIMENTS

### A. Polyp Phantoms

To thoroughly evaluate the performance and sensitivity of the proposed VTSB, we designed and fabricated 12 different types of CRC polyp phantoms (4 CRC polyp types, 3 different materials) mimicking the topology, size, and texture of CRC polyps based on realistic clinical cases using the Digital Anatomy Printer (J750, Stratasys, Ltd) [6]. These polyps replicate the pedunculated (type Ip), laterally spreading tumor (type LST), superficial elevate (type IIa), and depressed (type IIc) that represent the Paris classification [29]. Fig. 5 demonstrates, each unique polyp with its dimensions. We used materials with varying hardnesses as 00 45-60, A 30-40, and D 83-86 for Tissue Matrix/Agilus DM 400 (called M1), the mixture of Tissue Matrix and Agilus 30 Clear (called M2), and Vero Pure White (called M3), respectively [30].

### B. Experimental Setup

Fig. 4 demonstrates the experimental setup used to obtain the textural images of the proposed VTSB and measure the interaction forces between the sensor, and the custom-fabricated polyps. It consists of the VTSB, the fabricated CRC polyp phantoms, a transparent tube including randomly placed CRC polyps, a colonoscope (PENTAX EC 3840 L) for the demonstration of the applicability of the proposed VTSB, a single-row linear stage with 1 $\mu$m precision (M-UMR12.40, Newport), a digital force gauge with 0.02 N resolution (Mark-10 Series 5, Mark-10 Corporation) attached to the linear stage to precisely push the polyp phantoms and measure the applied interaction force to the VTSB gel layer, a Raspberry Pi 4 Model B for streaming and recording the video for the image processing algorithm, and the pneumatic actuation system (Programmable Air, Crowd Supply) for controlling the inflation and deflation of the VTSB as well as measuring the pressure inside the balloon to ensure the safety of the procedure. This pneumatic actuation system consists of two air pumps, three solenoid valves, and an air pressure sensor. We also utilized MESUR Lite basic data acquisition software (Mark-10 Corporation) to record the interaction forces between the VTSB and polyp phantoms.

### C. Experimental Procedure and Results

To perform experiments, VTSB was first fixed on a 3D printed rigid fixture on the optical table. Next, a polyp phantom was attached to the force gauge using the threaded connection printed at the base of each phantom. After preparing the hardware for measurements, both the Arduino Serial monitor, VTSB camera, and the Mesur Lite software were sequentially initialized. During the procedure, we recorded the pressure level inside the VTSB, the VTSB output images, as well as the synchronized interaction force values between the VTSB gel layer and the objects. Of note, these steps were repeated for all polyp types with three distinct materials (i.e., M1, M2, and M3). Fig. 5 summarizes the obtained images and corresponding interaction forces, and balloon pressures.

### D. Case Study- Colonoscopy Procedure

To validate the proposed working principle, described in Section II-A, and the performance of the VTSB while integrated with a colonoscope (PENTAX EC 3840 L) and working in a restricted environment, we also conducted a set of experiments inside a transparent round acrylic tube (Meccanixity) with dimensions of inner and outer diameter as 86 mm and 90 mm, respectively. To mimic a case with the randomized polyp existence in the colon, we attached 3 different types of fabricated polyp phantoms on the inner wall of the round tube, shown in Fig. 4. The inflation sequence of the VTSB is presented in Fig. 6 (a)-(f), clearly showing the successful operation of the VTSB inside the balloon. Of note, the exemplary obtained visual VTSB texture and corresponding pressures have already been shown in Fig. 5.

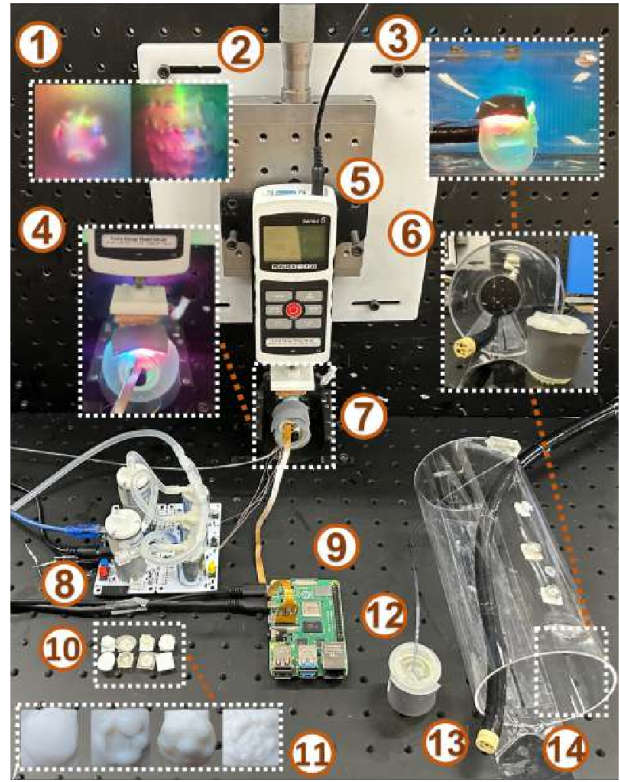

Fig. 4. Experimental Setup: ①- Image output of the VTSB sensor for both IIc CRC polyp, and the animal tissue sample, ②- M-UMR12.40 Precision Linear Stage, ③- A zoomed view of the VTSB in the transparent tube during Case Study 1, ④- A zoomed view of the VTSB interacting with an object, ⑤- Mark-10 Series 5 Digital Force Gauge, ⑥- A front view of the Case Study 1 setup, ⑦- Inflatable Vision-based Tactile Sensing Balloon, ⑧- Programmable Air, ⑨- Raspberry Pi 4, ⑩- CRC polyp phantoms, ⑪- A zoomed view of the CRC polyp phantoms, ⑫- An additional VTSB sensor, ⑬- PENTAX EC 3840 L colonoscope, ⑭- A transparent tube including randomly placed CRC polyps.

## IV. RESULTS AND DISCUSSION

Fig. 5 compares the visual outputs of the proposed novel VTSB sensor with a similar non-inflatable vision-based tactile sensor (i.e., GelSight (GS) sensor), developed in our previous study and tested on identical polyp phantoms [21]. This figure summarizes the visual output of VTSB tested on 4 different types of CRC polyp phantoms 3D printed with three different materials (i.e., M1, M2, and M3). The figure also shows the evolution of the textural outputs of each polyp under different interaction forces (between 0.3 N and 2.8 N) and corresponding pressure levels (between 6 kPa and 7.5 kPA). As can be clearly observed in this figure, VTSB exhibits considerably better performance than the GS sensor in creating high-quality images independent of the CRC polyps' type, texture, and stiffness, even at very low interaction forces (i.e. $\leq$ 1.7 N). Particularly, when the interaction force is less than 0.8 N, VTSB can provide high textural visual outputs as compared with the GS sensor. Of note, these high-quality images are obtained well below the safe interaction force (i.e., <13.5 N [24]) and under the safe pressure level (i.e., < 7.6 kPa [25]) for colonoscopy. Additionally, the performed case study clearly demonstrates the successful performance of the VTSB sensor in integration with a colonoscope and sequential mapping of the tube's internal surface. As can be observed in Fig. 6, thanks to the soft nature of the VTSB, this sensor is able to morph around the whole internal space of the tube comprising the attached CRC polyp phantoms, which creates an opportunity to map the high-quality textural details of the environment.

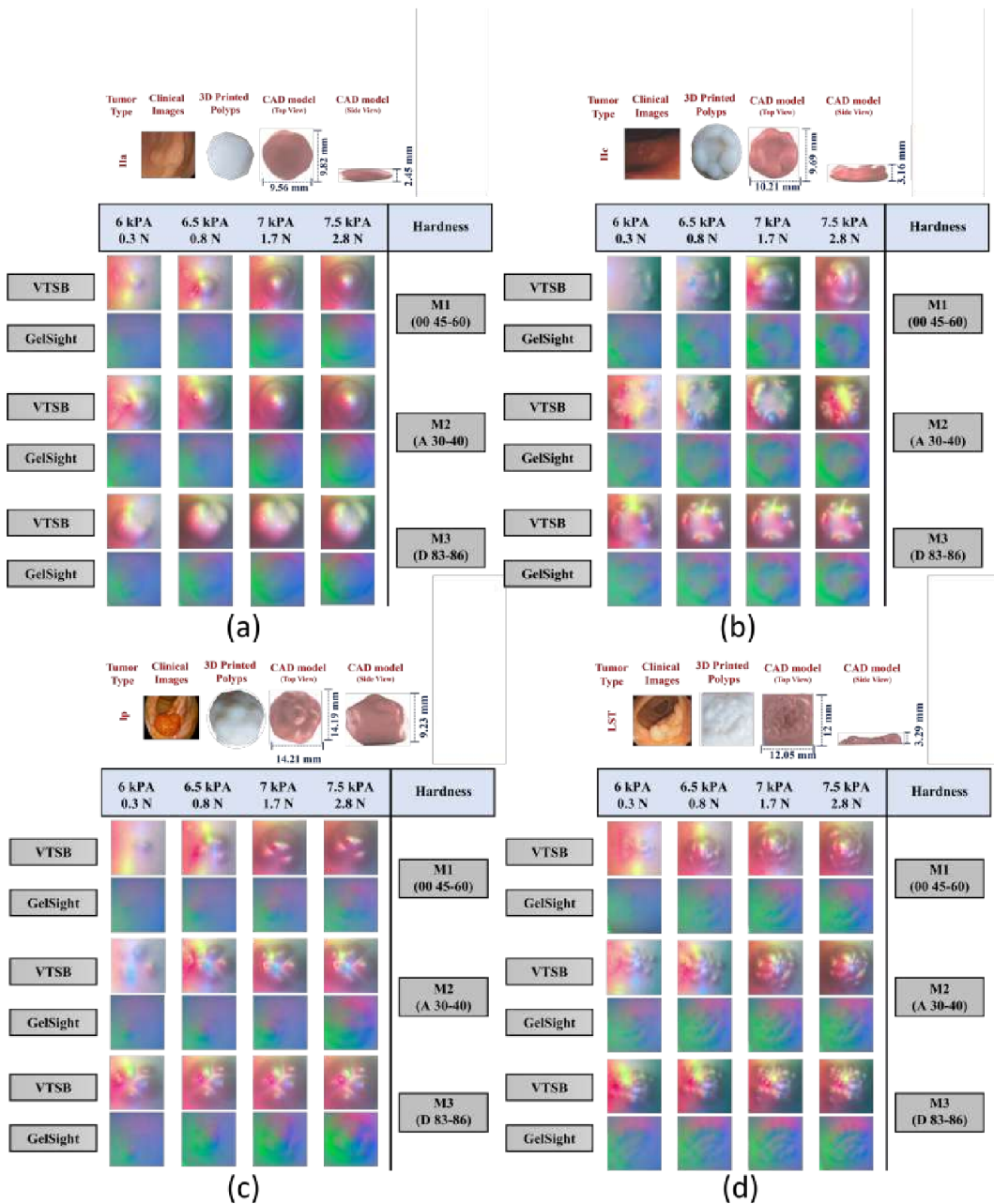

Fig. 5. Evolution of the textural visual outputs of VTSB sensor and its comparison with GS, at specified applied interaction force and internal pressure. Each block of the figure represents the outputs of VTSB for a unique CRC polyp made of three different materials (i.e., M1, M2, and M3). Of note, GS images were taken from the previous study [21], and pressure information does not apply to GS since it is a static sensor. (a) IIa, (b) IIc, (c) Ip, (d) LST.

## V. Conclusion and Future Work

In this study, we proposed a novel soft and inflatable VTSB that can enable (i) high-resolution, safe, and radiation-less imaging of the colon as well as (ii) sensitive and reliable identification of CRC polyps solely based on the textural images provided by the VTSB, to minimize the EDMR of CRC polyps and lower the risk of mortality in cancer patients. The performance of this novel sensor was thoroughly evaluated by performing various experiments on four different types of realistic 3D printed CRC polyps with three different material properties. To characterize the performance of the proposed VTSB, we measured the interaction force and the inflation pressure levels while obtaining the textural features of the polyp phantoms using VTSB. We also compared the performance of the VTSB with another similar non-inflatable vision-based tactile sensor while measuring the texture of identical phantoms. Results indicated that the proposed VTSB can obtain clear and high-fidelity images at a safe interaction force that is approximately 4 times less than the maximum allowable interaction force for a colonoscopy procedure and below the maximum allowable safe pressure of 7.6 kPa inside the colon. We also integrated the VTSB with the colonoscope and successfully demonstrated the workflow of the VTSB for a colonoscopic procedure.

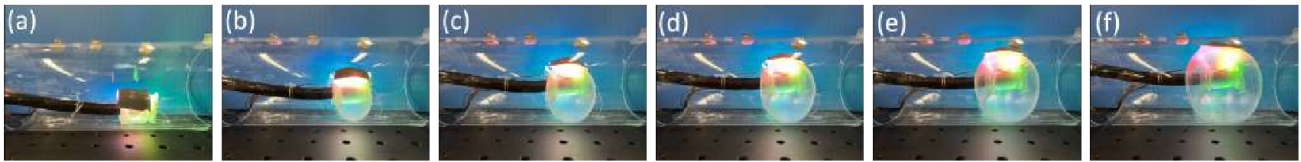

Fig. 6. Case Study: Attachment of VTSB to the colonoscope, showing the sequence of inflation and deflation.

## Acknowledgment

This research was supported by Basic Science Research Program through the National Research Foundation of Korea (NRF) funded by the Ministry of Education (2022R1A6A3A03063886).